\documentclass{rspublic}
\usepackage{epsfig}
\begin{document}

\title[Quantum information processing with trapped ions]{Quantum information processing\\  with trapped ions}

\author[D. J. Wineland and others]{D. J. Wineland, M. Barrett, J. Britton, J. Chiaverini, B.~DeMarco, W. M. Itano, B. Jelenkovi{\'c}$^{\dag}$,
C. Langer, D. Leibfried, V.~Meyer, T. Rosenband, and T.
Sch{\"a}tz}

\affiliation{Time and Frequency Division\\National Institute of Standards and Technology\\Boulder, CO, 80305-3328, USA\\
$\dag$ permanent address: Institute of Physics, Belgrade,
Yugoslavia}

\label{firstpage}

\maketitle

\begin{abstract}{quantum information processing, quantum computation, entanglement, trapped atomic ions}
Experiments directed towards the development of a quantum computer
based on trapped atomic ions are described briefly. We discuss the
implementation of single qubit operations and gates between
qubits. A geometric phase gate between two ion qubits is
described. Limitations of the trapped-ion method such as those
caused by Stark shifts and spontaneous emission are addressed.
Finally, we describe a strategy to realize a large-scale device.
\end{abstract}

\section{Introduction}

Efforts to realize experimentally the elements of quantum
computation (QC) using trapped atomic ions have been stimulated in
large part by a proposal by Cirac and Zoller (1995). In this
scheme, ions confined in a linear RF (Paul) trap are cooled and
form a stable spatial array whose motion is described by normal
modes. Two internal levels in each ion form a qubit (referred to
below as a spin qubit). The spacing between ions is typically
large enough ($> 1~ \mu m$) that the direct coupling of internal
states of ions is negligible, thereby precluding logic gates based
on internal-state interactions.  (An exception to this might be
``Dipole blockade" gates based on Rydberg transitions as
envisioned for neutral atoms (Jaksch {\em et al.} 2000), but these
gates are experimentally more challenging for ions because of the
higher energies between ground and Rydberg levels.)  To overcome
this limitation, Cirac and Zoller suggested cooling the ions to
their motional ground state and using the ground and first excited
state of a particular collective motional mode as a qubit (motion
qubit). The motional mode can act as a data bus to transfer
information between ions by mapping the spin qubit state of a
particular ion onto the selected motion qubit with a laser beam
focused onto that ion. The ability to construct a logic gate
between the motion qubit and another selected spin qubit, coupled
with the ability to perform single-spin qubit rotations provides
the basis for universal quantum computation (DiVincenzo 1995,
Barenco {\em et al.} 1995).

The ion-trap scheme satisfies the main requirements for a quantum
computer as outlined by DiVincenzo (2001): (1) a scalable system
of well-defined qubits, (2) a method to reliably initialize the
quantum system, (3) long coherence times, (4) existence of
universal gates, and (5) an efficient measurement scheme. Most of
these requirements have been demonstrated experimentally;
consequently, ion trap quantum processors are studied in several
laboratories. Here, we focus on experiments carried out at NIST
but note that similar work is being pursued at Aarhus, Almaden
(IBM), Hamburg, Hamilton (McMaster Univ., Ontario), Innsbruck, Los
Alamos (LANL), University of Michigan, Garching (MPI), Oxford, and
Teddington (NPL, U.K.).

\section{\bf Coherent control and entanglement}

The key entangling operation in the 1995 Cirac/Zoller scheme for
QC, and other schemes that rely on the ions' motion, is an
operation that couples a spin qubit with the motion in a
spin-state-dependent way. Assume that the spin qubit has a ground
state labeled $|\!\downarrow\rangle$ and a higher metastable state
labeled $|\!\uparrow\rangle$ that are separated in energy by
$\hbar \omega_0$. First assume that single-photon transitions
between these levels can be excited with a focused laser beam. For
simplicity, we will assume electric dipole transitions, but this
is easily adapted to other cases such as electric quadrupole
transitions as in the Innsbruck experiments (Roos {\em et al.}
1999). The interaction between an ion and the electric field of
the laser beam can be written as
\begin{equation}
H_I (t) = -\vec{d}\cdot\vec{E} = -\vec{d}\cdot E_0\hat{\epsilon}_L
\cos(k \widetilde{z} - \omega_L t + \phi),
\label{basic_interaction}
\end{equation}
where $\vec{d}$ is the electric dipole operator, $\widetilde{z}$
is the ion position operator for displacements from the ion's
equilibrium position (expanded in terms of normal mode operators),
$\hat{\epsilon}_L$ is the laser beam polarization, $k$ is the
laser beam's $k$-vector (taken to be parallel to $\hat{z}$, the
axis of the trap), $\omega_L$ is the laser frequency, and $\phi$
is the phase of the laser field at the mean position of the ion.
$E_0$ is the laser beam electric field amplitude at the ion, which
is assumed to be classical.  We characterize the laser field's
polarization with respect to a magnetic field $\vec{B}_0$ that
sets the quantization axis for the ions.  The polarization is
expressed in terms of left circular ($\hat{\sigma}_-$), linear
($\hat{\pi}$), and right circular ($\hat{\sigma}_+$) polarizations
so that $\hat{\epsilon}_L = e_-\hat{\sigma}_- + e_0\hat{\pi} +
e_+\hat{\sigma}_+$ with $|e_-|^2 + |e_0|^2 + |e_+|^2 = 1$. The
dipole operator $ \vec{d}$ is proportional to $\sigma^+ +
\sigma^-$, where $ \sigma^+ \equiv
|\!\uparrow\rangle\langle\downarrow\!|, \sigma^- \equiv
|\!\downarrow\rangle\langle\uparrow\!|$, and we take $\widetilde
{z} = z_0(a + a^\dag)$, where $a$ and $a^\dag$ are the lowering
and raising operators for the harmonic oscillator associated with
the selected motional mode (frequency $\omega_z$) and $z_0$ is the
extension of the ground-state wavefunction for the particular ion
(and mode) being addressed. We assume all other $z$ modes are
cooled to and remain in their ground states and for simplicity
have neglected them in $\widetilde{z}$. In the Lamb-Dicke limit,
where the extent of the ion's motion is much less than $\lambda /
2 \pi = 1/k$, we can write Eq. (\ref{basic_interaction}) (in the
interaction frame, and making the rotating wave approximation (see
for example, Wineland {\em et al.} 1998)) as
\begin{equation}
H_I \simeq  \hbar (\Omega e^{i\phi})\sigma^+ e^{-i(\omega_L -
\omega_0)t} [1 + i \eta (a e^{-i\omega_z t} + a^{\dagger}
e^{i\omega_z t})]+ h.c. \label{LD_limit_interaction}
\end{equation}
Here, $\Omega \equiv -E_0 \langle
\uparrow|\vec{d}\cdot\hat{\epsilon}_L|\!\downarrow \rangle /(2
\hbar) $ and $\eta \equiv k z_0$ is the Lamb-Dicke parameter ($\ll
1$ in the Lamb-Dicke limit).

For certain choices of $\omega_L$,\ $H_I$ is resonant and the spin
and motion can be coupled efficiently.  For example, when
$\omega_L = \omega_0 - \omega_z, H_I \simeq i \eta \hbar (\Omega
e^{i\phi}) \sigma^+ a + h.c.$. This is usually called the
``red-sideband" coupling and is formally equivalent to the
Jaynes-Cummings Hamiltonian from quantum optics (see, for example,
Raimond {\em et al.} 2001) . Here, $|\!\downarrow \rangle
\rightarrow |\!\uparrow \rangle$ transitions are accompanied by
$|n\rangle \rightarrow |n-1\rangle$ motional mode transitions.
When $\omega_L = \omega_0 + \omega_z$ (the blue sideband
frequency), $H_I \simeq i \eta \hbar (\Omega e^{i \phi})\sigma^+
a^{\dag}  + h.c.$, and $|\!\downarrow \rangle \rightarrow
|\!\uparrow \rangle$ transitions are accompanied by $|n\rangle
\rightarrow |n+1\rangle$ transitions. When $\omega_L = \omega_0$,
$H_I \simeq \hbar (\Omega e^{i \phi})\sigma^+  + h.c.$ and
$|\!\downarrow \rangle \rightarrow |\!\uparrow \rangle$
transitions do not change $n$. These ``carrier" transitions are
used to perform the single spin qubit rotations.

\subsection{Raman transitions}

Some experiments use two ground-state hyperfine levels as a qubit.
Coherent transitions between these levels can be implemented with
two laser beams that drive two-photon stimulated-Raman
transitions.  In this case, in Eq. (\ref{LD_limit_interaction}),
$k$ must be replaced by the difference $\Delta k = |\vec{k}_1 -
\vec{k}_2|$ between $k$-vectors for the two Raman beams (again
assumed to parallel to $\hat{z}$), $\omega_L$ and $\phi$ are
replaced by $\omega_1 -\omega_2$ and $\phi_1 - \phi_2$, the
frequency difference and phase difference between the laser beams
at the mean position of the ion, and $\Omega e^{i\phi}$ is
replaced by the resonant two-photon Raman Rabi rate
\begin{equation}
\Omega e^{i\phi} \leftrightarrow \frac{e^{i(\phi_1 - \phi_2)}}{4
\hbar^2} \sum_i  \langle \uparrow
|\vec{d}\cdot{E_2}\hat{\epsilon}_2|i \rangle\langle
i|\vec{d}\cdot{E_1}\hat{\epsilon}_1|\!\downarrow \rangle/\Delta_i.
\label{raman_rabi_gen}
\end{equation}
In this expression, the subscripts denote the two laser beams, $
|i \rangle$ are the (virtual, electronically excited) intermediate
states of the Raman process, and $\Delta_i$ are the detunings of
the Raman beams as indicated in Fig.~1. We have assumed $\Delta_i
\gg \gamma_i$ where $\gamma_i$ are the decay rates from the
intermediate states. With these substitutions, Eq.
(\ref{LD_limit_interaction}) applies in the Lamb-Dicke limit where
the Lamb-Dicke parameter is now given by $\eta \equiv \Delta k z_0
\ll 1$. For brevity, we will specialize to the
stimulated-Raman-transition case in what follows.
\begin{figure}
\centerline{\psfig{file=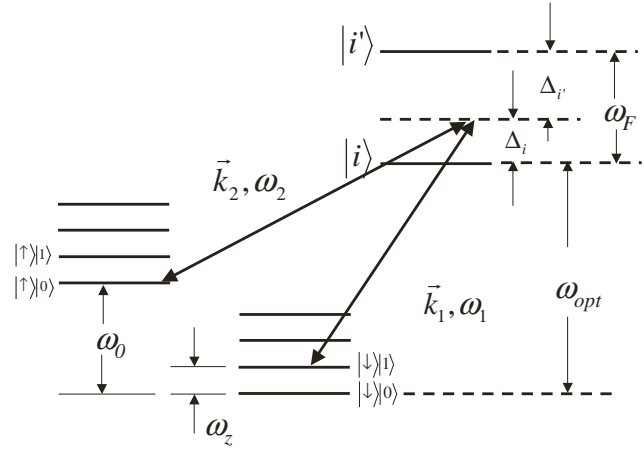, width=8cm}} 
\label{raman_transitions_gen} \caption{Schematic representation of
relevant energy levels for stimulated-Raman transitions (not to
scale). Shown are two ground-state hyperfine levels
($|\!\downarrow \rangle$ and $|\!\uparrow \rangle$) for one ion,
two (of possibly many) excited levels ($|i\rangle$ and
$|i'\rangle$), and the harmonic oscillator levels for one mode of
motion. Typically, $\omega_z \ll \omega_0 \ll \Delta_i,
\Delta_{i'}, \omega_F \ll \omega_{opt}$ where $\omega_{opt}$ is an
optical frequency.}
\end{figure}

When the difference frequency of the Raman beams is tuned to
resonance with the red or blue sidebands or carrier, the
interaction leads to the coherent evolution
\begin{equation}
|\!\downarrow \rangle|n \rangle \rightarrow
\cos(\Omega_{n,n'}t)|\! \downarrow \rangle|n \rangle -
ie^{i\phi}\sin(\Omega_{n,n'}t)|\! \uparrow \rangle|n' \rangle
\label{rabi_flopping_1}
\end{equation}
and
\begin{equation}\label{rabi_flopping_2}
|\!\uparrow \rangle|n \rangle \rightarrow
-ie^{-i\phi}\sin(\Omega_{n,n'}t)|\!\downarrow \rangle|n
\rangle+\cos(\Omega_{n,n'}t)|\!\uparrow \rangle|n'
 \rangle.
\end{equation}
When $n' = n \pm 1$ (blue or red sidebands), $\Omega_{n,n'} \equiv
\eta\Omega(n_>)^{1/2}$, where $n_>$ is the greater of $n$ or $n'$.
In Fig.~1, we show a red sideband transition for $n=1$.  When the
duration of this operation is adjusted to give a $\pi$ pulse, the
spin qubit to motion qubit mapping step required in the
Cirac/Zoller scheme $(\alpha|\!\downarrow \rangle +
\beta|\!\uparrow \rangle)|0\rangle \rightarrow |\!\downarrow
\rangle(\alpha|0\rangle + \beta|1\rangle)$ is executed. From
expressions (\ref{rabi_flopping_1}) and (\ref{rabi_flopping_2})
the entanglement between the spin and motion is evident since the
final state cannot in general be factored into the product of spin
and motional wavefunctions. Carrier transitions can also be
described by these expressions where $n = n'$ and $\Omega_{n,n'} =
\Omega$. For each ion we are free to choose $\phi = \phi_1 -
\phi_2 = 0$ but the phase of all operations on this ion must be
referenced to this choice.

In order to keep the duration of an entangling operation
relatively short, $\Omega_{n,n'}$ for $n \neq n'$ cannot be too
small relative to $\Omega_{n,n}$.  Therefore $\eta$ must be chosen
large enough that our interaction does not rigorously satisfy the
Lamb-Dicke criterion and correction factors must be added to the
expressions for the Rabi frequencies $\Omega_{n,n'}$. These
correction factors, which can be called Debye-Waller factors
(Wineland {\em et al.} 1998), have been observed by Meekhof {\em
et al.} (1996) and also form the basis for a controlled-not (CNOT)
gate between motion and spin (Monroe {\em et al.} 1997, DeMarco
{\em et al.} 2002).

\subsection{Rabi frequencies}

For convenience of notation, in the case $\omega_1 - \omega_2
\simeq \omega_0$, we will designate $\omega_1 = \omega_b$ the
``blue" Raman beam frequency and $\omega_2 = \omega_r$ as the
``red" Raman beam frequency.  Similarly, we let $E_1
\hat{\epsilon}_1 = E_b \hat{\epsilon_b} = E_b (b_-\hat{\sigma}_- +
b_0\hat{\pi} + b_+\hat{\sigma}_+)$ and $E_2 \hat{\epsilon}_2 = E_r
\hat{\epsilon_r} = E_r (r_-\hat{\sigma}_- + r_0\hat{\pi} +
r_+\hat{\sigma}_+)$.  As an example relevant to the NIST
experiments, we consider Raman transitions between the $2s\
^2S_{1/2}$ hyperfine states $|F=2,m_F=2\rangle \equiv
{|\!\downarrow\rangle}$ and $|F=1,m_F=1\rangle \equiv |\!\uparrow
\rangle$ in $^9$Be$^+$ where we couple through the ion's $2p$ fine
structure levels. (In this example, $|i\rangle \equiv 2p\ \
\!\!^2P_{1/2}$ and $|i'\rangle \equiv 2p\ \ \!\!^2P_{3/2}$ in
Fig.~1.) Using the appropriate Clebsch-Gordon coefficients to
evaluate the terms $\langle \downarrow,\uparrow|\vec{d}\cdot{E_j}
\hat{\epsilon}_j|i \rangle$ in Eq. (\ref{raman_rabi_gen}), we find
the carrier Rabi frequency to be
\begin{equation}
\Omega_{\downarrow,\uparrow} = \frac{g_bg_r}{\sqrt{6}}
\bigl[b_0r_+ +
b_-r_0\bigr]\frac{\omega_F}{\Delta(\Delta-\omega_F)},\label{rabi1}
\end{equation}
where $\Delta$ is the detuning of the Raman beams from the
$^2P_{1/2}$ level ($\Delta_i$ in the figure), $\omega_F/2 \pi \
(\simeq 198$ GHz) is the fine structure splitting, $\omega_0/2\pi
\ \simeq 1.25$ GHz, and $g_{b,r} \equiv E_{b,r}|\langle ^2P_{3/2},
F=3, m_F=3|\vec{d}\cdot \hat{\sigma}_+|\!\downarrow \rangle|/2
\hbar$.

In some cases we will also find it useful to adjust the Raman beam
difference frequency to a multiple of the mode oscillation
frequency.  When $\omega_b - \omega_r \simeq \omega_z$, the
interaction (in the Lamb-Dicke limit) can be written
\begin{equation}
H_I(m_S) \simeq  \hbar \Omega(m_S)
 e^{-i(\omega_b - \omega_r)t} [1 + i \eta (a e^{-i\omega_z t} + a^{\dagger}
e^{i\omega_z t})]+ h.c. \label{displacement}
\end{equation}
where
\begin{equation}
\Omega(m_S) \equiv \frac{e^{i(\phi_b - \phi_r)}}{4 \hbar^2}
\sum_{i} \langle m_S |\vec{d}\cdot{E_r}\hat{\epsilon}_r|i
\rangle\langle i|\vec{d}\cdot{E_b}\hat{\epsilon}_b|m_S
\rangle/\Delta_i \label{displacement_rabi}
\end{equation}
with $m_S \in \{\downarrow, \uparrow \}$.   Assuming $\omega_b -
\omega_r = \omega_z$, and approximating Eq. (\ref{displacement})
with the resonant term, an oscillating, spatially uniform force is
created that can act as a displacement operator (Meekhof {\em et
al.} 1996). It takes the form (Wineland {\em et al.} 1998)
\begin{equation}
D(t) = D(\eta \Omega(m_S) t) = e^{(\eta \Omega(m_S) t)a^{\dag} -
(\eta \Omega(m_S) t)^* a} .
\end{equation}
When $\Omega(\downarrow) \neq \Omega(\uparrow)$, we can create
logic gates as described in the example below.

\subsection{Stark shifts}

In the above discussion we have neglected Stark shifts of the
levels caused by the non-resonant laser beam electric fields. In
the limit that $\Delta_i \gg \gamma$, the Stark shift from the
$j$th beam on the $m_S$ level is given by
\begin{equation}\label{Stark_shifts}
  \delta(m_S,j) = \frac{1}{4 \hbar^2} \sum_i |\langle m_S|\vec{d}\cdot{E_j}
  \hat{\epsilon}_j|i \rangle|^2/\Delta_i.
\end{equation}
In general, both levels $|\!\downarrow \rangle$ and $|\!\uparrow
\rangle$ are shifted by both laser beams.  For the $|2,2\rangle =
|\!\downarrow \rangle \leftrightarrow |1,1\rangle = |\!\uparrow
\rangle$ $^9$Be$^+$ Raman transition example, the net Stark shift
is
\begin{eqnarray}\label{Stark_beryllium}
\delta(\uparrow) - \delta(\downarrow) =  g_b^2
\Biggl[\frac{(\frac{1}{6}b_-^2 + \frac{1}{3}b_0^2 +
\frac{1}{2}b_+^2)}{\Delta + \omega_0} + \frac{(\frac{5}{6}b_-^2 +
\frac{2}{3}b_0^2 + \frac{1}{2} b_+^2)}{\Delta-\omega_F +
\omega_0}\nonumber
\\ - \frac{(\frac{2}{3}b_-^2 + \frac{1}{3}b_0^2)}{\Delta}
-\frac{(\frac{1}{3}b_-^2 +\frac{2}{3}b_0^2 + b_+^2)}{\Delta -
\omega_F}\Biggr] + g_r^2 \Biggl[ \frac{(\frac{1}{6}r_-^2 +
\frac{1}{3}r_0^2 + \frac{1}{2}r_+^2)}{\Delta}\nonumber
\\ + \frac{(\frac{5}{6}r_-^2 + \frac{2}{3}r_0^2 +
\frac{1}{2}r_+^2)}{\Delta-\omega_F} - \frac{(\frac{2}{3}r_-^2 +
\frac{1}{3}r_0^2)}{\Delta-\omega_0} - \frac{(\frac{1}{3} r_-^2 +
\frac{2}{3}r_0^2 + r_+^2)}{\Delta - \omega_0 - \omega_F} \Biggr].
\label{22-11Stark}
\end{eqnarray}
In Eq. (\ref{22-11Stark}) we have assumed that the (quantizing)
magnetic field is small enough that Zeeman shifts can be neglected
and that the $^2P_{1/2}$ and $^2P_{3/2}$ hyperfine splittings are
negligible comparable to $\Delta_i$ (The hyperfine splitting of
the $^2P_{1/2}$ level is approximately 237 MHz and that of the
$^2P_{3/2}$ level is less than 1 MHz.). In the limit that we can
also assume $\omega_0 \ll \Delta, \omega_F$, Eq.
(\ref{22-11Stark}) reduces to
\begin{equation}
\delta(\uparrow) - \delta(\downarrow) \simeq \bigr[g_b^2(b_-^2 -
b_+^2) +g_r^2(r_-^2 - r_+^2)\bigl] \Biggl[\frac{\omega_F}{2
\Delta(\Delta-\omega_F)}\Biggr].
\end{equation}
In this approximation, we see that the Stark shifts vanish if we
use linearly polarized light for the Raman beams.  Moreover, if we
use the exact expression for $\delta(\uparrow) -
\delta(\downarrow)$, we find that the Stark shift can be tuned to
zero by adjusting $r_- - r_+$ and/or $b_- - b_+$ to small non-zero
values.

\subsection{Spontaneous emission}

A fundamental limitation to the coherence of atomic qubits is
spontaneous emission.  When the qubits are formed from
ground-state hyperfine levels, memory is not affected by
spontaneous emission since these levels have very long radiative
lifetimes. The problem arises from spontaneous emission during
Raman transitions (see for example Plenio and Knight 1997, Di
Fidio and Vogel 2000, Budini {\em et al.} 2002). As an estimate of
the decoherence rate due to spontaneous emission we calculate the
total spontaneous emission rate $R_{SE}$ from the $2p$ levels.
This rate is equal to the sum of the probabilities $P_i$ that each
intermediate (excited) state is occupied times its decay rate
$\gamma_i$:
\begin{equation}
R_{SE} = \sum_i \gamma_i P_i = \frac{1}{4 \hbar^2} \sum_i
\sum_{\{j=r,b\}} \sum_{\{m_S=\downarrow,\uparrow\}} P_{m_S}
\gamma_i|\langle m_S|\vec{d} \cdot E_j
\hat{\epsilon}_j|i\rangle|^2/\Delta_i^2,
\end{equation}
where $P_{m_S}$ is the probability of being in the $m_S$ ground
state. For our $^9$Be$^+$ example, we find
\begin{eqnarray}
R_{SE} = \gamma P_{\downarrow}\Biggl[\frac{g_b^2(\frac{2}{3}b_-^2
+ \frac{1}{3}b_0^2)}{\Delta^2} + \frac{g_b^2(\frac{1}{3}b_-^2 +
\frac{2}{3}b_0^2 + b_+^2)}{(\Delta - \omega_F)^2} +
\frac{g_r^2(\frac{2}{3}r_-^2 + \frac{1}{3}r_0^2)}{(\Delta-\omega_0)^2}\nonumber \\
+ \frac{g_r^2(\frac{1}{3} r_-^2 + \frac{2}{3}r_0^2 +
r_+^2)}{(\Delta - \omega_0 - \omega_F)^2}\Biggl] + \gamma
P_{\uparrow}\Biggl[\frac{g_b^2(\frac{1}{6}b_-^2 + \frac{1}{3}b_0^2
+ \frac{1}{2}b_+^2)}{(\Delta + \omega_0)^2}\ \ \ \ \ \ \ \nonumber \\
+ \frac{g_b^2(\frac{5}{6}b_-^2 + \frac{2}{3}b_0^2 + \frac{1}{2}
b_+^2)}{(\Delta-\omega_F + \omega_0)^2} +
\frac{g_r^2(\frac{1}{6}r_-^2 + \frac{1}{3}r_0^2 +
\frac{1}{2}r_+^2)}{\Delta^2} + \frac{g_r^2(\frac{5}{6}r_-^2 +
\frac{2}{3}r_0^2 + \frac{1}{2}r_+^2)}{(\Delta-\omega_F)^2}\Biggl],
\label{SE}
\end{eqnarray}
where $\gamma/2\pi = 19.4$ MHz. Of course, we want to minimize the
probability of spontaneous emission during qubit operations.  As
one measure of this, we can calculate the probability of
spontaneous emission during the time $\tau_{\pi}$ required to
carry out a $\pi$ pulse on the carrier transition, given by
$\Omega \tau_{\pi} = \pi/2$ (Eqs. (\ref{rabi_flopping_1}) and
(\ref{rabi_flopping_2})).  Since we also want to suppress Stark
shifts, we take $b_- \simeq b_+$ and $r_- \simeq r_+$. During the
$\pi$ transition, $\langle P_{\downarrow} \rangle = \langle
P_{\uparrow} \rangle = 1/2$. With these assumptions, Eq.
(\ref{SE}) gives the probability of spontaneous emission during a
carrier $\pi$ pulse as
\begin{equation}
P_{SE} = R_{SE} \tau_{\pi} \simeq \frac{\pi \gamma}{6
|\Omega_{\downarrow,\uparrow}|} \bigl(g_b^2 + g_r^2 \bigr)
\Biggl[\frac{1}{\Delta^2} + \frac{2}{(\Delta - \omega_F)^2}
\Biggr].
\end{equation}
This expression is minimized with the choices $\Delta = (\sqrt{2}
- 1) \omega_F \simeq 82$ GHz, $g_b = g_r$, and $b_0 = 1, r_+ = r_-
= 1/\sqrt{2}$ ($b_+ = b_- = r_0 = 0$) giving the value $P_{SE} =
(8 \pi/\sqrt{6})(\gamma/\omega_F) \simeq 0.001$. This probability
will be increased by $1/\eta$ for sideband transitions (in the
Lamb-Dicke limit). For this reason, $^9$Be$^+$ will not be a good
choice for the ultimate qubit since probabilities for errors
during a gate operation must be on the order of $10^{-4}$ or
smaller to be able to incorporate error correction into long
computations (Steane 2002).

To suppress the effects of spontaneous emission, we want an ion
with a small ratio of spontaneous decay rate $\gamma$ to fine
structure splitting (see for example, Wineland 2002).  To make a
straightforward comparison between different ions, we calculate
the probability of spontaneous emission for various ions with half
odd-integer nuclear spin $I$ for a $\pi$ pulse on the carrier of
the $|F=I-1/2,m_F=0\rangle \leftrightarrow |F=I+1/2,m_F=0\rangle$
transition.  The Rabi rate for these transitions will be
independent of $I$.  Moreover, the frequencies of these
transitions deviate from their values at magnetic field $\vec{B} =
0$ by an amount proportional to $|\vec{B}|^2$ and are therefore
first-order independent of field fluctuations as $|\vec{B}|
\rightarrow 0$.  Again making the approximations $\omega_0,\gamma
\ll \omega_F,|\Delta|$, we find
\begin{equation}
\Omega_{0 \leftrightarrow 0} = \frac{g_b g_r \omega_F}{3
\Delta(\Delta - \omega_F)}(b_-r_- -b_+r_+) \label{rabi0-0}
\end{equation}
and
\begin{equation}
R_{SE} = \gamma \frac{g_b^2 + g_r^2}{3} \Biggl[\frac{1}{\Delta^2}
+ \frac{2}{(\Delta - \omega_F)^2} \Biggr].\label{SE_2}
\end{equation}
For simplicity we have taken the value of $\gamma$ to be that
corresponding to the relevant $^2P_{1/2}$ level, we have assumed
pure $^2P_{1/2}$ and $^2P_{3/2}$ configurations, and we have
included coupling to only this pair of fine-structure levels.
These approximations should be most accurate for the lighter ions.
We can minimize $R_{SE}/|\Omega_{0 \leftrightarrow 0}|$ by
choosing $\Delta = (\sqrt{2}-1)\omega_F$, $g_b = g_r$, and the
experimentally convenient choice of orthogonal linear
polarizations for the Raman beams ($b_- = r_- = b_+ = -r_+ =
1/\sqrt{2}$). This leads to a probability of spontaneous emission
during a carrier $\pi$ pulse given by $P_{SE} = 2 \sqrt{2} \pi
\gamma/\omega_F$. The Stark shift of the transition is equal to
\begin{equation}
\delta_{0 \leftrightarrow 0} = -\frac{(g_b^2 + g_r^2)\omega_0}{3}
\Biggl[\frac{1}{\Delta^2} + \frac{2}{(\Delta-\omega_F)^2} \Biggr],
\end{equation}
which is independent of polarization.  For the conditions above,
we find  $\delta_{0 \leftrightarrow 0} = -4\sqrt{2}|\Omega_{0
\leftrightarrow 0}|\omega_0/\omega_F$. In Table~1 we tabulate
$|\delta_{0 \leftrightarrow 0}/\Omega_{0 \leftrightarrow 0}|$ and
$P_{SE}$ for a few ions of interest.
\begin{table}
\label{P_SE} \caption{Probability of spontaneous emission $P_{SE}$
during a two-photon stimulated-Raman $\pi$ pulse on the carrier of
the $^2S_{1/2}, |F=I-1/2,m_F=0\rangle \leftrightarrow
|F=I+1/2,m_F=0\rangle$ transition for various ions of interest in
quantum computing. ($I$ is the nuclear spin, $\nu_F \equiv
\omega_F/2\pi$, $\nu_0 \equiv \omega_0/2\pi$.)}
\begin{tabular}{ccccccc}
\hline
ion & $I$ & $\gamma/2\pi(\mathrm{MHz})$ & $\nu_F(\mathrm{THz})$ & $\nu_0(\mathrm{GHz})$ & $|\delta_{0 \leftrightarrow 0}/\Omega_{0 \leftrightarrow 0}|$ & $P_{SE}$\\
\hline
$^9$Be$^+$ & 3/2 & 19.4 & 0.198 & 1.25 & $3.6\times10^{-2}$ & $8.7\times10^{-4}$\\
$^{25}$Mg$^+$ & 5/2 & 43 & 2.75 & 1.79 & $3.6\times10^{-3}$ & $1.4\times10^{-4}$\\
$^{43}$Ca$^+$ & 7/2 & 22.4 & 6.7 & 3.26 & $2.8\times10^{-3}$ & $3.0\times10^{-5}$\\
$^{67}$Zn$^+$ & 5/2 & 76 & 26.2 & 7.2 & $1.6\times10^{-3}$ & $2.6\times10^{-5}$\\
$^{87}$Sr$^+$ & 9/2 & 21.7 & 24 & 5.00 & $1.2\times10^{-3}$ & $8.0\times10^{-6}$\\
$^{113}$Cd$^+$ & 1/2 & 44.2 & 74 & 15.2 & $1.2\times10^{-3}$ & $5.3\times10^{-6}$\\
$^{199}$Hg$^+$ & 1/2 & 54.7 & 274 & 40.5 & $8.4\times10^{-4}$ & $1.8\times10^{-6}$\\
\hline
\end{tabular}
\end{table}

In the Lamb-Dicke limit, sideband transitions and two-qubit gates
will necessarily have a higher probability of spontaneous emission
since the Rabi frequencies for these processes scale as the
carrier frequency times $\eta$, whereas the spontaneous emission
rate will remain unchanged. Therefore, from Table~1 we see that,
in order to suppress the probability of spontaneous emission
during Raman transitions, a heavy ion such as Cd$^+$ (Blinov {\em
et al.} 2002) will ultimately be required. Moreover, even for
heavy ions, there is not much headroom on fidelity to reach fault
tolerance. Therefore, for fault-tolerant quantum computation,
values of $\eta$ should not be too small compared to unity; that
is, the Lamb-Dicke limit will not be a good approximation. This
implies the need for precise control of the motion (e.g., ground
state cooling) since the gate Rabi rates will depend on the
motional states.  In addition, schemes that incorporate error
resistant methods such as adiabatic passage, spin echo, and
composite pulses like those used in NMR (Cummings {\em et al.}
2002, Jones 2002) will have to be used sparingly, if at all, since
these techniques increase the probability of spontaneous emission
for each logical operation.

In principle, we can relax these restrictions by making $|\Delta|
\gg \omega_{FS}$.  In the above discussion we have estimated
decoherence using the total rate of spontaneous emission (Eqs.
(\ref{SE}) and (\ref{SE_2})).  However, when $|\Delta| \gg
\omega_{FS}$, spontaneous emission is dominated by Raleigh
scattering and the decoherence rate will be suppressed compared to
the total spontaneous emission rate (see, for example, Cline {\em
et al.} (1994)).  The cost for making $|\Delta| \gg \omega_{FS}$
is that the Raman laser beam intensities must increase as
$\Delta^2$ to maintain the same Rabi frequency. Moreover, for
large values of $\omega_{FS}$, we must consider the decohering
effects of spontaneous Raman scattering from levels other than the
two fine-structure levels considered here.

Finally, we note that the spontaneous emission problem could be
essentially eliminated by driving single-photon hyperfine carrier
and sideband transitions at frequencies $\omega_0$ and $\omega_0
\pm \omega_z$. However, it appears that in practice, currently
attainable field gradients (necessary for sideband transitions)
would lead to relatively small values of $\eta$ and therefore slow
sideband transition Rabi rates (see, for example Wineland {\em et
al.} 1992 and Mintert and Wunderlich 2001).

\subsection{State initialization and detection}

To initialize the qubits for each experiment, we use a combination
of internal-state optical pumping to pump to the
$|\!\downarrow\rangle$ state and laser cooling to optically pump
the motional modes to their ground states (Monroe {\em et al.}
1995a, King {\em et al.} 1998, Roos {\em et al.} 1999). As in many
atomic physics experiments, the observable in the ion trap
experiments is the ion's spin qubit state. We can efficiently
distinguish $|\!\downarrow \rangle$ from $|\!\uparrow \rangle$
using a cycling transition to implement ``quantum jump" detection
(Blatt \& Zoller 1988).

\section{\bf Gates}

The previous sections summarize the basic sources of entanglement
in ion experiments from which universal logic gates have been
constructed. For example, a CNOT and $\pi$-phase gate between the
motion and spin qubit for a single ion has been realized by Monroe
{\em et al.} (1995b), Wineland {\em et al.} (1998), and DeMarco
{\em et al.} (2002). Also, using the scheme suggested by
S{\o}rensen and M{\o}lmer (1999, 2000) and Solano {\em et al.}
(1999), the NIST group realized a universal gate between two spin
qubits (Sackett {\em et al.} 2000, Kielpinski {\em et al.} 2001).
Compared to the original Cirac and Zoller gate (1995), this last
gate has the practical advantages that (1) laser-beam focusing
(for individual ion addressing) is not required, (2) it can be
carried out in one step, (3) it does not require use of an
additional internal state, and (4) it does not require precise
control of the motional state (as long as the Lamb-Dicke limit is
satisfied). From this gate, a CNOT gate can be constructed
(S{\o}rensen and M{\o}lmer (1999)). Below, we discuss a ($\pi$)
phase gate based on a spin-dependent displacement operator.

\subsection{Geometrical phase gate}

We have recently realized a ($\pi$) phase gate between two spin
qubits (Leibfried {\em et al.} 2002) that carried out the
transformations: $|\!\downarrow \rangle|\! \downarrow \rangle
\rightarrow |\!\downarrow \rangle|\!\downarrow \rangle$,
$|\!\downarrow \rangle| \!\uparrow \rangle \rightarrow |
\!\downarrow \rangle|\! \uparrow \rangle$, $| \!\uparrow
\rangle|\! \downarrow \rangle \rightarrow |\! \uparrow \rangle|\!
\downarrow \rangle$, and $|\! \uparrow \rangle|\! \uparrow \rangle
\rightarrow -|\! \uparrow \rangle|\! \uparrow \rangle$.  Consider
constructing a closed path in phase space (for a particular
motional mode) so that the state returns to its original position.
We can derive the effects of this transformation by constructing
the closed path from a series of successive applications of the
displacement operator (see for example, Walls \& Milburn 1994)
with infinitesimal displacements. The net effect is that the wave
function describing the ion (or ions) acquires an overall phase
shift that depends on the area enclosed by the path.

The second element required for the gate is to make the area of
the path be spin-dependent.  This is accomplished by making the
displacement in phase space with a spin-dependent optical dipole
force.  This, at first, seems to be impossible if we require that
the net Stark shift between the $|\!\downarrow \rangle$ and
$|\!\uparrow \rangle$ states be equal to zero since optical dipole
forces and Stark shifts are proportional. However, we can make the
time-averaged Stark shift $\langle \delta(\uparrow) -
\delta(\downarrow) \rangle$ be equal to zero for durations much
greater than $2 \pi/\omega_z$ and still realize a spin dependent
displacement operator if we choose the polarizations of the Raman
laser beams appropriately. The basic idea is that we make the
dipole force (sinusoidally varying at frequency $\omega_z$) have a
different phase for the $|\!\downarrow \rangle$ and $|\!\uparrow
\rangle$ states.

\begin{figure}
\centerline{\psfig{file=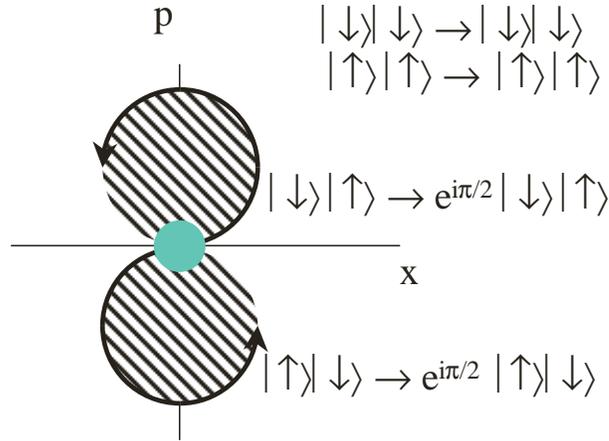, width=8cm}} 
\caption{ Schematic representation of the displacements of the
axial stretch-mode amplitude in phase space for the four basis
states of the two spin qubits.  The detuning and amplitude of the
displacements are chosen to give a $\pi/2$ phase shift on the
$|\!\downarrow \rangle|\!\uparrow \rangle$ and $|\!\uparrow
\rangle|\!\downarrow \rangle$ states while the $|\!\downarrow
\rangle|\!\downarrow \rangle$ and $|\!\uparrow \rangle|\!\uparrow
\rangle$ states are unaffected because the optical dipole forces
for these states do not couple to the stretch
mode.}\label{Didi_gate}
\end{figure}

For simplicity, assume both Raman beams are linearly polarized
with polarization perpendicular to the quantizing magnetic field.
Also assume that the red-Raman beam polarization has an angle
$\kappa$ with respect to that of the blue-Raman beam. Hence we
have $E_b \hat{\epsilon_b} = E_b (b_-\hat{\sigma}_- +
b_+\hat{\sigma}_+)$ and $E_r \hat{\epsilon_r} = E_r
(r_-\hat{\sigma}_- + r_+\hat{\sigma}_+ e^{i 2 \kappa})$ where, as
before, we take $b_- \simeq b_+ \equiv b$ and $r_- \simeq r_+
\equiv r$ to make $\langle \delta(\uparrow) - \delta(\downarrow)
\rangle = 0$. For our $^9$Be$^+$ example, the interaction
Hamiltonian is given by Eqs. (\ref{displacement}) and
(\ref{displacement_rabi}), with
\begin{equation}
\Omega(\downarrow) = g_r g_b b r\Biggl[\frac{2/3}{\Delta} +
\frac{1/3 + e^{i 2 \kappa}}{\Delta - \omega_F} \Biggr], \ \ \
\Omega(\uparrow) = g_r g_b b r\Biggl[\frac{1/6 + 1/2\
e^{i2\kappa}}{\Delta} + \frac{5/6 + 1/2\
e^{2i\kappa}}{\Delta-\omega_F}\Biggr].
\end{equation}
By choosing the polarizations orthogonal ($\kappa = \pi/2$) and
$\Delta = (\sqrt{2} - 1)\omega_F$, we find
\begin{equation}
\Omega(\downarrow) =  \frac{2 g_b g_r b r}{3(3 \sqrt{2}
-4)\omega_F} = - 2 \Omega(\uparrow). \label{dipole_force}
\end{equation}
Hence, the dipole force on the $|\!\downarrow \rangle$ state is
twice that on the $|\!\uparrow \rangle$ state and in the opposite
direction.  (This same displacement operator has been used
previously to create Schr{\"o}dinger cat states of a single ion
(Myatt {\em et al.} 2000).)  To implement this gate on two ions,
the Raman transition beams were separated in frequency by
$\sqrt{3}\omega_z + \delta $, where $\sqrt{3}\omega_z$ is the
stretch mode frequency for two ions aligned along the $z$ axis and
$\delta$ is a small detuning ($\ll \omega_z$). The separation of
the ions was adjusted to be an integer multiple of $2\pi/\Delta k$
so that the optical-dipole force on each ion was in the same
direction if the ions were in the same spin state but in opposite
directions if the spin states was different (Eq.
(\ref{dipole_force})). This had the effect that the application of
the laser beams to the $| \!\downarrow \rangle|\! \uparrow
\rangle$ and $|\! \uparrow \rangle| \!\downarrow \rangle$ states
caused excitation on the stretch mode but the motion was not
excited when the ions were in the $|\! \downarrow \rangle|\!
\downarrow \rangle$ or $|\! \uparrow \rangle| \uparrow \rangle$
states. The detuning $\delta$ and duration of the displacement
pulses ($2 \pi/\delta$) were chosen to make one complete
(circular) path in phase space with an area that gave a phase
shift of $\pi/2$ on the $|\!\downarrow \rangle|\!\uparrow \rangle$
and $|\!\uparrow \rangle|\!\downarrow \rangle$ states (Fig.
\ref{Didi_gate}). Under these conditions, the overall
transformation was: $|\!\downarrow \rangle|\!\downarrow
\rangle\rightarrow|\!\downarrow \rangle|\!\downarrow \rangle$,
$|\!\downarrow \rangle|\!\uparrow \rangle\rightarrow
e^{i\pi/2}|\!\downarrow \rangle|\!\uparrow \rangle$, $|\!\uparrow
\rangle|\!\downarrow \rangle\rightarrow e^{i\pi/2} |\!\uparrow
\rangle|\!\downarrow \rangle$, and$|\!\uparrow \rangle|\!\uparrow
\rangle\rightarrow|\!\uparrow \rangle|\!\uparrow \rangle =
e^{i\pi} e^{-i\pi} |\!\uparrow \rangle|\!\uparrow \rangle$.
Therefore, this operator acts like the product of an operator that
applies a $\pi/2$ phase shift to the $|\!\uparrow \rangle$ state
on each ion separately (a non-entangling gate) and $\pi$ phase
gate between the two ions. The $\pi/2$ phase shifts can be removed
by applying additional single qubit rotations or accounted for in
software. (In an algorithm carried out by a series of single-qubit
rotations and two-qubit phase gates, the extra phase shifts can be
removed by appropriately shifting the phase of subsequent or prior
single-qubit rotations.)

This phase gate is a particular case of the more general formalism
developed by Milburn {\em et al.} (1999), S{\o}rensen and
M{\o}lmer (1999), and Wang {\em et al.} (2001). Compared to the
original Cirac and Zoller gate (1995) this gate shares the same
advantages as the S{\o}rensen and M{\o}lmer gate (2000) as well as
some additional technical advantages (Leibfried {\em et al.}
2002).

\section{Whither quantum computation?}

Although simple operations among a few ion qubits have been
demonstrated, a viable quantum computer must look towards scaling
to very large numbers of qubits. As the number of ions in a trap
increases, several difficulties are encountered. For example, the
addition of each ion adds three vibrational modes. It soon becomes
nearly impossible to spectrally isolate the desired vibrational
mode unless the speed of operations is slowed to undesirable
levels (see for example Wineland {\em et al.} 1998, Steane and
Lucas, 2000). Furthermore, since error correction will most likely
be incorporated into any large processor, it will be desirable to
measure and reset ancilla qubits without disturbing the coherence
of logical qubits.  Since ion qubits are typically measured by
means of state-dependent laser scattering, the scattered light
from ancilla qubits held in a common trap may disturb the
coherence of the logical qubits.

For these and other reasons, it appears that a scalable ion-trap
system must incorporate arrays of interconnected traps, each
holding a small number of ions. The information carriers between
traps might be photons (Cirac {\em et al.} 1997, Pellizzari 1997,
De Voe 1998), or ions that are moved between traps in the array.
In the latter case, a ``head" ion held in a movable trap could
carry the information by moving from site-to-site as in the
proposal of Cirac and Zoller (2000). Similarly, as has been
proposed at NIST, we could shuttle ions around in an array of
interconnected traps (Wineland {\em et al.} 1998, Kielpinski {\em
et al.} 2002). In this last scheme, the idea is to move ions
between nodes in the array by applying time-dependent potentials
to ``control" electrode segments.  To perform logic operations
between selected ions, these ions are transferred into an
``accumulator" trap for the gate operation. Before the gate
operation is performed, it may be necessary to sympathetically
re-cool the qubit ions with another ion species. Subsequently,
these ions are moved to memory locations or other accumulators.
This strategy always maintains a relatively small number of
motional modes that must be considered and minimizes the problems
of ion/laser-beam addressing using focused laser beams. Such
arrays also enable highly parallel processing and ancilla qubit
readout in separate trapping regions so that the logical ions are
shielded from the scattered laser light.  Some of the initial
steps towards this scheme have been reported by Rowe {\em et al.}
(2002).

The obstacles to building a large-scale quantum computer appear to
be technical rather than fundamental. However since it will be a
long time before a useful processor is constructed, it is helpful
to have some intermediate goals upon which projections about
scaling can be made.  Examples goals might be the demonstration of
repetitive error correction (see for example, Nielsen and Chuang
2000) and the realization of gates at the fault-tolerant level of
fidelity (Steane 2002).

\section{Acknowledgements} \noindent We thank Marie Jensen and Windell Oskay for helpful comments on the
manuscript.  This work was supported by the U. S. National
Security Agency (NSA) and Advanced Research and Development
Activity (ARDA) under Contract No. MOD-7171.00, and the U. S.
Office of Naval Research (ONR).  The article is a contribution of
NIST and not subject to U.S. copyright.

\end{document}